\documentclass[11pt,oneside]{article}
\usepackage{academicons, lastpage, booktabs}
\usepackage[dvipsnames]{xcolor}
\usepackage[english]{babel}
\usepackage{amsfonts,amssymb,amsmath,amsthm}
\usepackage[absolute,overlay]{textpos}
\usepackage{float}
\usepackage{fancyhdr, graphicx}
\usepackage{enumerate}
\usepackage[margin=1in]{geometry}
\usepackage{tikz}
\usepackage{layout}
\usepackage{titlesec}
\usepackage{titling}
\usepackage{authblk}
\usepackage{xcolor, colortbl}
\usepackage{geometry}
\usepackage{siunitx} 
\usepackage{hyperref}
\usepackage{eso-pic}
\usepackage{marginnote}
\usepackage{doi}

\usepackage{tabularray}
\usepackage{csquotes}
\usepackage[section]{placeins}
\usepackage{gensymb}

\definecolor{Gallery}{rgb}{0.937,0.937,0.937}
\definecolor{Zumthor}{rgb}{0.925,0.956,1}
\definecolor{LemonChiffon}{rgb}{1,1,0.78}
\definecolor{SnowyMint}{rgb}{0.807,1,0.803}
\definecolor{Azalea}{rgb}{0.984,0.878,0.874}

\definecolor{coolblack}{rgb}{0.0, 0.18, 0.30}
\definecolor{carmine}{rgb}{0.59, 0.0, 0.09}
\definecolor{myblue}{rgb}{0.0, 0.48, 0.65}
\definecolor{cadetgrey}{rgb}{0.57, 0.64, 0.69}
\definecolor{lightgray}{gray}{0.9}

\hypersetup{
    colorlinks=true,
    linkcolor=coolblack,
   filecolor=magenta,      
    urlcolor=cyan,
    citecolor=coolblack,
    pdfhighlight=/P,
                 }
                                  
\usepackage{libertinus}
\usepackage[T1]{fontenc}
\usepackage{orcidlink} 

\usepackage{caption}
\usepackage{fancyhdr}

\titleformat*{\section}{\Large\bfseries\color{black}}
\titleformat*{\subsection}{\large\bfseries\color{black}}
\titleformat*{\subsubsection}{\bfseries\color{black}}

\makeatletter
\renewcommand\@makefntext[1]{\leftskip=0em\hskip0em\@makefnmark#1}
\makeatother
\newcommand\blfootnote[1]{%
  \begingroup
  \renewcommand\thefootnote{}\footnote{#1}%
  \addtocounter{footnote}{-1}%
  \endgroup
}

\providecommand{\keywords}[1]
{
  \small	
  \textbf{\textsf{Keywords---}} #1
}

\renewcommand{\tableautorefname}

\newcommand{\mycomment}[1]{} 
\newcommand*\mytitle{Numerical simulations of a RF-RF hybrid plasma torch with argon at atmospheric pressure}
\newcommand*\myauthor{Loann Terraz \textit{et al.}} 
\newcommand*\mycorrmail{loann.terraz@kit.edu} 
\newcommand*\myDOI{10.46298/ops.16373}
\newcommand*\myarxiv{2508.13858}
\newcommand*\myHAL{-}
\newcommand*\myzenodo{-} 
\newcommand*\myvolume{ICPIG 2025}
\newcommand*\myyear{2026} 
\newcommand*\mynumber{1 } 
\newcommand*\mydaterec{August 20, 2025}
\newcommand*\mydaterevised{October 8, 2025} 
\newcommand*\mydateaccepted{January 6, 2026} 
\newcommand*\mydatepublished{January 14, 2026}

\newlength{\myshift}
\usepackage[firstpage=true]{background}
\backgroundsetup{contents={\includegraphics[width=1.\paperwidth]{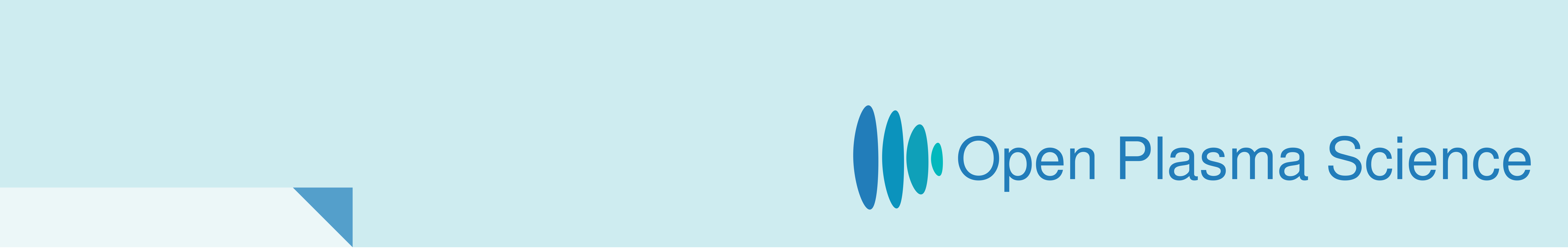}},scale=1,placement=top,opacity=1}

\pretitle{\vspace{-00pt}      \begin{flushleft}   \Large  \color{black} \selectfont  } 
     
\title{\textbf{\mytitle}}

\posttitle{\par \end{flushleft} } 

\preauthor{\vspace{-00pt} \begin{flushleft}  \color{black} \selectfont} 

\author[,1]{Loann Terraz\orcidlink{0000-0002-6735-4255}$^{*}$}   
\author[1]{Biruk T. Alemu\orcidlink{0009-0001-5420-6464}}
\author[1]{Santiago Eizaguirre\orcidlink{0000-0001-6076-1483}}

\affil[1]{\small Light Technology Institute, Karlsruhe Institute of Technology, Karlsruhe, Germany}

\postauthor{\par\end{flushleft}} 

\date{}

\begin{document}

\maketitle
\thispagestyle{empty}
\blfootnote{$^*$ Corresponding author: \textsf{\href{\mycorrmail}{\mycorrmail}}}
\blfootnote{Cite as: \myauthor, \mytitle, \textit{Open Plasma Science}  \myvolume , \mynumber (\myyear), doi: \myDOI}


\reversemarginpar

\marginnote{\begin{flushleft}
\sf \bf \footnotesize \color{coolblack} History
\end{flushleft}}[-\myshift]
\addtolength{\myshift}{-0.5cm} 
\marginnote{
\begin{flushleft} \footnotesize
Received \mydaterec
\end{flushleft}}[-\myshift]

\addtolength{\myshift}{-0.5cm} 
\marginnote{
\begin{flushleft} \footnotesize
Revised \mydaterevised
\end{flushleft}}[-\myshift]

\addtolength{\myshift}{-0.5cm} 
\marginnote{
\begin{flushleft} \footnotesize
Accepted \mydateaccepted
\end{flushleft}}[-\myshift]

\addtolength{\myshift}{-0.5cm} 
\marginnote{
\begin{flushleft} \footnotesize
Published \mydatepublished
\end{flushleft}}[-\myshift]

\addtolength{\myshift}{-1cm} 

\marginnote{\begin{flushleft}
\sf \bf \footnotesize \color{coolblack} Identifiers
\end{flushleft}}[-\myshift]
\addtolength{\myshift}{-0.5cm} 
\marginnote{
\begin{flushleft} \footnotesize
DOI \href{https://doi.org/\myDOI}{\myDOI} 
\end{flushleft}}[-\myshift]
\addtolength{\myshift}{-0.5cm} 
\marginnote{
\begin{flushleft} \footnotesize
HAL \href{https://hal.science/\myHAL}{\myHAL} 
\end{flushleft}}[-\myshift]
\addtolength{\myshift}{-0.5cm} 
\marginnote{
\begin{flushleft} \footnotesize
ArXiv \href{https://arxiv.org/\myarxiv}{\myarxiv} 
\end{flushleft}}[-\myshift]
\addtolength{\myshift}{-0.5cm} 
\marginnote{
\begin{flushleft} \footnotesize
Zenodo \href{https://https://zenodo.org/records/\myzenodo}{\myzenodo} 
\end{flushleft}}[-\myshift]

\addtolength{\myshift}{-1cm} 
\marginnote{\begin{flushleft}
\sf \bf \footnotesize \color{coolblack} Supplementary Material
\end{flushleft}}[-\myshift]
\addtolength{\myshift}{-0.5cm} 
\marginnote{
\begin{flushleft} \footnotesize
- 
\end{flushleft}}[-\myshift]

\addtolength{\myshift}{-1cm} 
\marginnote{\begin{flushleft}
		\sf \bf \footnotesize \color{coolblack} Licence
\end{flushleft}}[-\myshift]
\addtolength{\myshift}{-0.5cm}
\marginnote{
	\begin{flushleft} 
		\href{https://creativecommons.org/licenses/by/4.0/}{\includegraphics[width=0.55\marginparwidth]{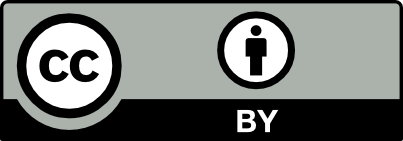}}
\end{flushleft}}[-\myshift]
\addtolength{\myshift}{-1.25cm} 
\marginnote{
	\begin{flushleft} \footnotesize
		\copyright  The Authors
\end{flushleft}}[-\myshift]

\addtolength{\myshift}{11.75cm} 
\marginnote{\begin{flushleft}
\sf \footnotesize \color{coolblack} \textsc{\textbf{Vol. \myvolume, No. \mynumber (\myyear)}}
\end{flushleft}}[-\myshift]


\addtolength{\myshift}{-1.5cm} 
  
\marginnote{\begin{flushleft}
\sf \normalsize \color{coolblack} \textsc{\textbf{Special issue}}
\end{flushleft}}[-\myshift]

\begin{flushleft}
\textbf{\textsf{Abstract}}
\vspace{5pt} 
\end{flushleft}

 We report numerical results regarding the minimum sustaining coil excitation current for a RF-RF hybrid torch operating at two different frequencies. The first coil is excited at a high-frequency, while the second coil is set at a medium frequency. The filling gas is argon, at atmospheric pressure. We use the modeling software COMSOL Multiphysics\textsuperscript{\textregistered}~to describe the evolution of key parameters when: (i) the distance between the two coils changes, (ii) the power of the high frequency coil changes. We discuss the radial temperature profiles, the axial velocities and the heat convected at the end of the medium-frequency coil. The latter is compared with the total heat conduction to the plasma confinement tube wall.


\vspace{20pt} 


\begin{flushleft}

\keywords{RF-RF hybrid torch, ICP, dual frequency, thermal plasma}

\end{flushleft}

\newpage

\newgeometry{ left=20mm, right=20mm,  bottom=3.5cm }

\pagestyle{fancy}
\setlength{\headheight}{15pt}
\fancyhead{} 
\fancyhead[L]{\footnotesize{\href{https://doi.org/\myDOI}{\myDOI}}} 
\fancyhead[R]{\footnotesize{\myauthor}}

\fancyfoot{} 
\fancyfoot[R]{\color{coolblack}{ \thepage \hspace{1pt} | \pageref{LastPage}}}
\fancyfoot[C]{\footnotesize{\color{black} Open Plasma Science \textbf{\myvolume}, No.\mynumber (\myyear)}}

\renewcommand{\headrulewidth}{0.4pt}
\renewcommand{\footrulewidth}{0.4pt}
\renewcommand{\footruleskip}{5pt}
\renewcommand\footrule{\hrule width\textwidth}
\renewcommand\headrule{\hrule width\textwidth}


\tableofcontents


\section{Introduction}
\label{sec:intro}

Inductively coupled plasma (ICP) at atmospheric pressure is an electricity-based alternative to the classic gas-based combustion in the context of industrial heating applications, thanks to its high output enthalpy. Compared to electrode-based plasmas, it has the advantage of a longer lifetime and is contamination-free. However, operating such plasma may prove difficult and requires careful characterization of numerous factors' influence, one of them being the frequency of the inductive current. In general, the higher the frequency the lower the minimum power required to ignite and sustain the plasma (cf. p.~857-859 in \cite{boulos2023} for instance), but the cost of the corresponding electronics also increases substantially. This issue is partially solved by using two coils: one at high frequency (HF) to assist the ignition and operation of the torch, and one at medium frequency (MF) to further couple power into the plasma without increasing the overall costs drastically. It also has the advantage of increasing plasma stability. This design is referred to as RF-RF plasma torch, tandem plasma, or RF-hybrid plasma.

While single-coil ICPs are extensively described in the literature, publications dealing with RF-RF ICPs are more scarce. The idea is not new though, starting with Floyd \textit{et al.} \cite{floyd1966} in 1966, followed by Allen \textit{et al.} \cite{allen1987} in 1987, then by Uesugi \textit{et al.} \cite{uesugi1988} in 1988, and Kameyama \textit{et al.} \cite{kameyama1990} in 1990. Kuraishi \textit{et al.} \cite{kuraishi2013} developed a RF-RF system with modulation of the coil current in 2013, with a focus on nanoparticle synthesis. This work was further developed through numerical models by Siregar \textit{et al.} \cite{siregar2019}, and Onda \textit{et al.} \cite{onda2020}. Other authors also investigated hybrid torches, with a focus on modeling~\cite{bernardi2004,frolov2019}, and diamond growth~\cite{zuo2017,li2018}.

The final goal, in the context of this work, is to couple about $1~MW$ into the plasma, which would be feasible with HF electronics but not desirable cost-wise. On the other hand, using only a MF coil proved to be difficult in terms of ignition and operation. Hence a hybrid solution where the HF coil is igniting the plasma at a reasonably low power and where the MF coil provides the rest of the power in a customable and cheaper way. The work presented here is the first step towards this final goal. We want to quantify to what extent sustaining the plasma with one HF coil and one MF coil is easier than with the MF coil alone. To do so, we focus on parameters related to the plasma operation: the minimum sustaining current (MSI) and its corresponding power. We study their evolution when two parameters of the setup are modified: (i) the distance between the two coils, and (ii) the power of the HF coil. We investigate the range of possible values for electronics quantities, namely the coil power, voltage, current, and impedance. Once the torch is fully characterized, these parameters can be tuned during operation in order to reach the desired convective heat at the torch output, while keeping the conductive heat towards walls below a certain threshold value.

\section{Modeling the hybrid torch}
\label{sec:model}

\subsection{Torch schematics}

\begin{figure}[!h]
	\centering
	\includegraphics[height=13.25cm]{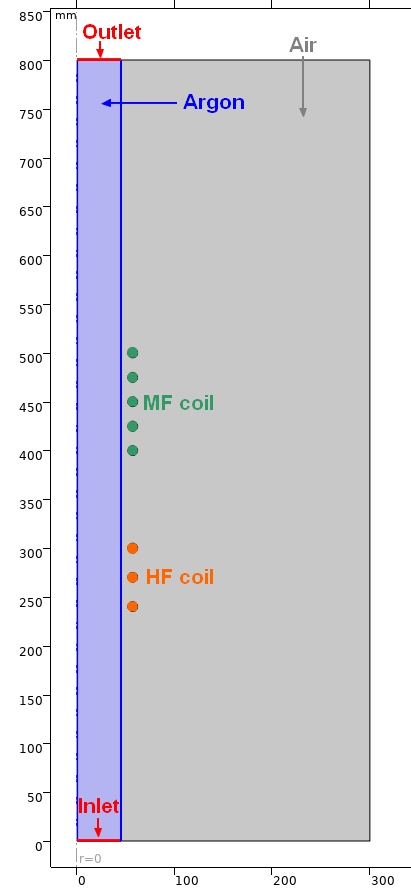}
	\caption{2D axisymmetric diagram of the plasma torch.}
	\label{fig:torch_schematics}
\end{figure}

The geometry, depicted in figure~\ref{fig:torch_schematics}, consists in:
\begin{itemize}
	\item A burner, i.e., the cylindrical torch and its wall + inlet + outlet, of length $L_{\rm burner} = 800~mm$ and radius $R_{\rm burner} = 45~mm$. The volume defined by the burner boundaries is called `Plasma domain' hereafter. The filling gas is argon. 
	\item A high-frequency coil: the copper helix (3D) of $N_{\rm HF} = 3$~turns is modeled as $3$ disks belonging to the same coil group. It has a length $L_{\rm HF} = 70~mm$, with a radial gap of $7~mm$ from the burner wall and starting at the axial distance $z_{\rm HF} = 235~mm$. The radius of the copper tube is $R_{\rm coil} = 5~mm$.
	\item A medium-frequency coil: the copper helix (3D) of $N_{\rm MF} = 5$~turns is modeled as $5$ disks belonging to the same coil group. It has a length $L_{\rm MF} = 110~mm$, with a radial gap of $7~mm$ from the burner wall and starting at the axial distance $z_{\rm MF} = z_{\rm HF} + L_{\rm HF} + D_{\rm coil}$, where $D_{\rm coil}$ is the distance between the coils and is equal to $100~mm$ by default. The radius of the copper tube is $R_{\rm coil} = 5~mm$.
	\item The rest is defined as ambient air at atmospheric pressure and $293~K$, over a volume large enough to ensure that most of the magnetic lines generated by the coils are fitting in. 
\end{itemize}

\subsection{Implementation and simplifications}
This study sets the basis for a future, more developed model of a RF-RF ICP torch. As such, we use the following assumptions:
\begin{itemize}
	\item The filling gas is pure argon, at a pressure $p = 1~atm$.
	\item The plasma is assumed to be in local thermodynamic equilibrium, and is described by macroscopic quantities like thermal and electrical conductivities (no kinetic reactions implemented).
	\item The plasma is electrically neutral.
	\item We use 2D axisymmetric simulations, which drastically reduce the computational cost compared with 3D simulations. 
	\item Radiation is not considered. Only conduction and convection are taken into account for the heat transfers, which may result in a rather large overestimation of the heat convected at the outlet. Some typical ratios of conduction, convection and radiation can be found p. 27 in \cite{thorpe1969} for instance. Another consequence is the shape of the temperature profiles, which typically exhibit a maximum at $r > 0~mm$ when radiation is included (cf. figure~7 in \cite{mensing1969}).
	\item The burner wall is modeled as a boundary condition, for the sake of computational simplicity. It is set to room temperature: $293.15~K$, depicting an ideal cooling system.
\end{itemize}

\subsection{Main equations}

In COMSOL Multiphysics\textsuperscript{\textregistered}, the \textit{Equilibrium Inductively Coupled Plasma} model couples the following interfaces:
\begin{enumerate}
	\item The \textit{Magnetic Fields} interface, solving Maxwell's equations in the whole computational domain:
	\begin{equation}
		\nabla \times \mathbf{H} = \mathbf{J},
	\end{equation}
	
	\begin{equation}
		\mathbf{B} = \nabla \times \mathbf{A},
	\end{equation}
	
	\begin{equation}
		\mathbf{J} = \sigma \mathbf{E} + j\omega \mathbf{D} + \sigma \mathbf{u}\times \mathbf{B} + \mathbf{J_e},
	\end{equation}
	
	\begin{equation}
		\mathbf{E}= -j\omega \mathbf{A},
	\end{equation}
	where $\mathbf{H}$ is the magnetic field intensity, $\mathbf{J}$ is the current density, $\mathbf{B}$ is the magnetic field, $\mathbf{A}$ is the magnetic vector potential, $\sigma$ is the electrical conductivity, $\mathbf{E}$ is the electric field, $j$ is the complex number such as $j^2=-1$, $\omega$ is the angular radio frequency, $\mathbf{D}$ is the displacement current, $\mathbf{u}$ is the fluid velocity, and $\mathbf{J_e}$ is the coil current. Using the vacuum dielectric constant $\varepsilon_0$ and the vacuum permeability $\mu_0$, we have the relations:
	\begin{equation}
		\mathbf{D} = \varepsilon_0 \mathbf{E},
	\end{equation}
	
	\begin{equation}
		\mathbf{B} = \mu_0 \mathbf{H}.
	\end{equation}
	Within the plasma volume, the term $\mathbf{J}$ simplifies to:
	\begin{equation}
		\mathbf{J} = \sigma(\mathbf{E} + \mathbf{u} \times \mathbf{B}).
	\end{equation}
	
	We used two different magnetic interfaces, one for each coil, with the principle of superposition: each magnetic field of different frequency is treated separately.

	\item The \textit{Heat Transfer in Fluids} interface, modeling the heat transfer by conduction and convection in the plasma volume:
	
	\begin{equation}
		\label{eq:ht}
		\rho C_p \frac{\partial T}{\partial t} + \rho C_p \mathbf{u} \cdot \nabla T + \nabla \cdot \mathbf{q} = Q + Q_p + Q_{vp},
	\end{equation}
	
	\begin{equation}
		\mathbf{q} = -k \nabla T,
	\end{equation}
	where $\rho$ is the density of the gas, $C_p$ is the specific heat at constant pressure, $T$ is the temperature, $\mathbf{q}$ is the heat flux by conduction and $k$ is the thermal conductivity. The 3 terms on the right-hand side of eq. \eqref{eq:ht} are the heat sources, with $Q_p$ the work done by pressure change and $Q_{vp}$ the viscous dissipation. The last term $Q$ accounts for resistive heating, volumetric net radiation loss if included (not the case in this work) and enthalpy transport. In this model, it is calculated as:
	\begin{equation}
		Q = \mathbf{E} \cdot \mathbf{J} + \frac{\partial}{\partial T} \left( \frac{5k_B T}{2q} \right) (\nabla T \cdot \mathbf{J} ),
	\end{equation}
	where $k_B$ is the Boltzmann constant and $q$ the electric charge. 
	
	\item The \textit{Laminar Flow} interface, solving the Navier-Stokes equations for conservation of momentum and conservation of mass. We assume a weakly compressible flow and we neglect gravity effects. Equations are solved within the plasma volume and, in the magnetohydrodynamics context, are expressed as:
	
	\begin{equation}
		\rho \frac{\partial \mathbf{u}}{\partial t} + \rho( \mathbf{u} \cdot \nabla) \mathbf{u} = \nabla \cdot \left[ -p \mathbf{I} + \mathbf{K} \right] + \mathbf{F},
	\end{equation}
	
	\begin{equation}
		\frac{\partial \rho}{\partial t} + \nabla \cdot (\rho \mathbf{u} )= 0,
	\end{equation}
	where $p$ is the pressure, $\mathbf{I}$ is the unit matrix, $\mathbf{K} = \mu \left( \nabla \mathbf{u} + (\nabla \mathbf{u})^T \right)$ is the viscous stress, $\mu$ being the dynamic viscosity, and $\mathbf{F}$ is a volume force, which is the Lorentz force in our ICP study:
	\begin{equation}
		\mathbf{F} = \frac{1}{2} \Re \left( \mathbf{J} \times \mathbf{B^*} \right).
	\end{equation}
	
\end{enumerate}

\subsection{Default parameters}
Otherwise stated, all the parameters depicted in table~\ref{tab:defaultValues} have the same value in every simulation. The current is given in peak values, as opposed to root-mean-square values.

\begin{table}[!ht]
	\centering
	\caption{Default values used for the different simulations.}
	\label{tab:defaultValues}
	\begin{tblr}{
			cells = {LemonChiffon},
			row{1} = {Gallery},
			row{2} = {Zumthor},
			row{3} = {Zumthor},
			row{6} = {SnowyMint},
			row{7} = {Azalea},
			row{8} = {Azalea},
		}
		\textbf{Variable}       & \textbf{Value} & \textbf{Description}           \\
		$f_{\rm HF}$         & 13.56 MHz      & HF coil frequency              \\
		$P_{\rm HF}$         & 3 kW           & HF coil excitation power       \\
		$f_{\rm MF}$         & 200 kHz        & MF coil frequency              \\
		$I_{\rm MF}$         & 500 A          & MF coil excitation current     \\
		$D_{\rm coil}$       & 100 mm         & Distance between the two coils \\
		$Q_{\rm inlet}$      & 100 L/min      & Argon flow at the inlet        \\
		$\alpha_{\rm swirl}$ & 0$\degree$     & Swirl angle of the gas inflow  
	\end{tblr}
\end{table}

\subsection{Mesh construction}
\label{subsec:meshconstruction}

\begin{figure}[!ht]
	\centering
	\includegraphics[width=1\linewidth]{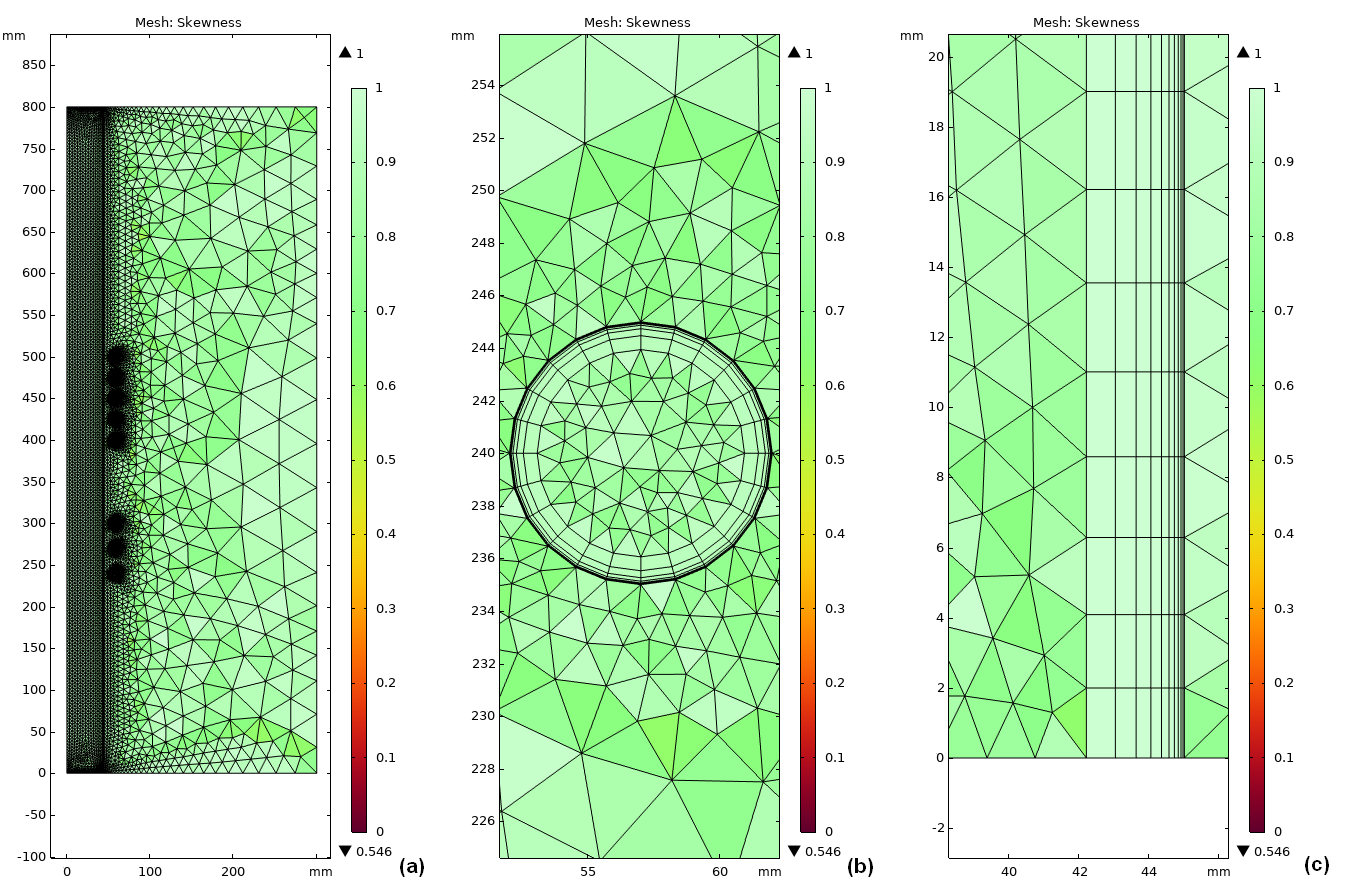}
	\caption{Default mesh skewness: (a) global view, (b) zoom on coil grid and (c) zoom on boundary layers at the wall.}
	\label{fig:default_mesh}
\end{figure}

We first use COMSOL to automatically generate a mesh, with the size criteria ``fine'', then rework it around the critical parts. The first boundary layer within the coil is set to $10^{-6}~mm$, with a stretching factor of $2$ for a total of $20$ layers. Similarly, the first boundary layer of the quartz wall is set to $4\times10^{-2}~mm$, with a stretching factor of $1.4$, for a total of $10$ layers. Due to the absence of inlet walls, we do not need corner refinement in this model. Finally, the inlet and outlet edges are set to a maximum element size of $2~mm$ to ensure a proper flow description. The resulting mesh, shown in figure~\ref{fig:default_mesh}, consists in a total of 18,596 elements. Using the skewness as the default quality measure, its average element quality is $0.8917$, a value of $1$ being ideal. Although some coarser mesh would allow solver convergence for the default case from table~\ref{tab:defaultValues}, it was not always true for the extreme conditions of our parametric study. Consequently, we found this grid to be a good compromise between quality of the results and computational speed, and it was used for all the simulations presented in this work. The sensitivity of the results to the mesh element size is discussed in subsection~\ref{subsec:gridinpendence}.

\section{Evolution of the minimum sustaining current}
\label{sec:results}

All the results presented in this section were obtained with transient studies, i.e., time-resolved simulations where equations are solved in the frequency domain. The operating conditions (current, gas flow, temperature at the boundaries) were kept constant long enough to reach a stable state. In this report, the main quantity of interest is the MSI. Although the notion of minimum sustaining power (MSP) is more often used, we decided to simulate the MF coil excitation with current rather than power, as it is closer to how we operate our lab facilities.  

A characteristic of ICPs is the presence of hysteresis: the transition from E-mode (capacitive coupling) to H-mode (inductive coupling) occurs at a higher RF current while ramping up than the H-E transition while ramping down. Note that the E-mode is equivalent to plasma extinction in our model. To obtain converging simulations, we had to set the initial gas, wall and inlet temperatures at $8000~K$.  Consequently, the corresponding current values for the ignition are not relevant and the hysteresis will not be characterized here. Only the MSI obtained while ramping down the current is presented in this section, as it is obtained with proper wall and inlet temperatures of $293.15~K$. The reason to do so is twofold:
\begin{itemize}
	\item Numerical instabilities can prevent convergence. It often occurs when the temperature gradient is too great.
	\item Coupling electromagnetic power into the plasma requires a non-negligible electrical conductivity $\sigma$. By default in COMSOL, $\sigma$ is implemented as a material property function of temperature only, with $\sigma_{\rm Ar}(T < 4500~K) < 1~S/m$. It implies that, except for extremely large (and non-realistic) values of current, the ignition will not take place at ambient temperature in our simulations. While experimenting in the laboratory, some free charge carriers are always present in the gas, whether they come from scarce kinetic ionization, radiation or directly generated from a sparkling device. These free charge carriers are not handled in the model.   
\end{itemize}

As a preliminary step, it is of interest to calculate: (i) the MSP of the HF coil when the MF coil is not working, (ii) the MSI of the MF coil when the HF coil is not working. We obtained the following results:
\begin{itemize}
	\item HF coil alone: MSP $\in [990~W ; 1050~W]$, for a corresponding coil current of $89.4~A$.
	\item MF coil alone: MSI $= 444~A$, for a corresponding coil power of $15.2~kW$.
\end{itemize}
As expected, sustaining a plasma at high frequencies requires drastically less power/current than at medium or low frequencies. However, as mentioned previously, the costs of the corresponding electronics also greatly increase with the frequency when one wants to couple more power.   

\begin{figure}[!ht]
	\centering
	\includegraphics[width=0.7\linewidth]{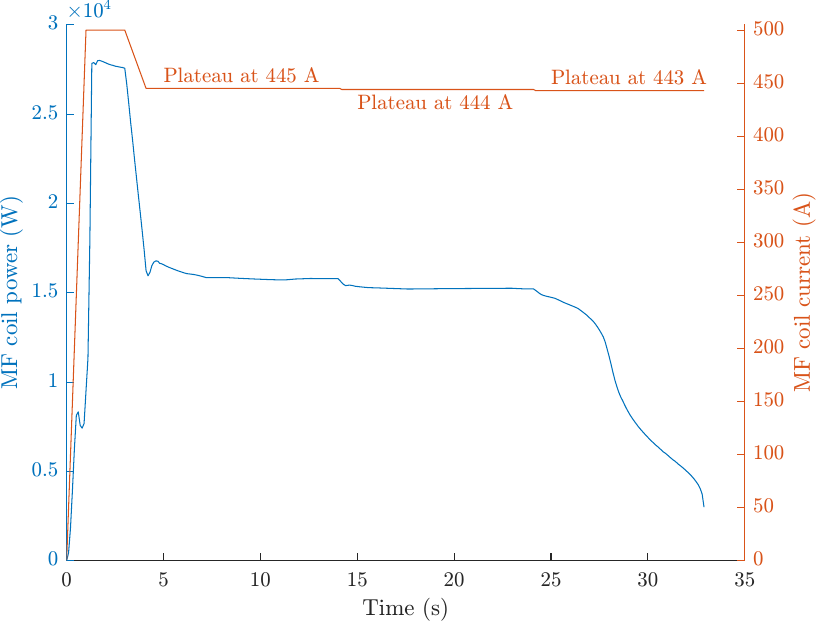}
	\caption{Typical current curve with plateau, and its corresponding power, to find an accurate value of the MSI. The MSI is defined as the last current value giving a constant coil power. In this case, only the MF coil is on.}
	\label{fig:MSI_MF_coil_alone}
\end{figure}

To find the MSI of the MF coil, the same strategy was adopted in all the simulations (cf. figure~\ref{fig:MSI_MF_coil_alone}): 
\begin{enumerate}
	\item Its excitation current is first ramped up in $\sim1~s$. Meanwhile, the inlet and wall temperatures are ramped down from $8000~K$ to $293.15~K$. The inflow volume rate is ramped up from $0~L.min^{-1}$ to $Q_{\rm inlet}$.
	\item Then, the current is set constant for a few seconds, to reach a stable state.
	\item Finally, the current is slowly ramped down with inclusion of plateau at the estimated MSI values. It is worth mentioning that a too fast decrease in current may lead to plasma extinction at higher values than the MSI. 
\end{enumerate}

\subsection{Varying the distance between the coils}
\label{subsec:coilDistance}
\begin{figure}[!ht]
	\centering
	\includegraphics[width=0.7\linewidth]{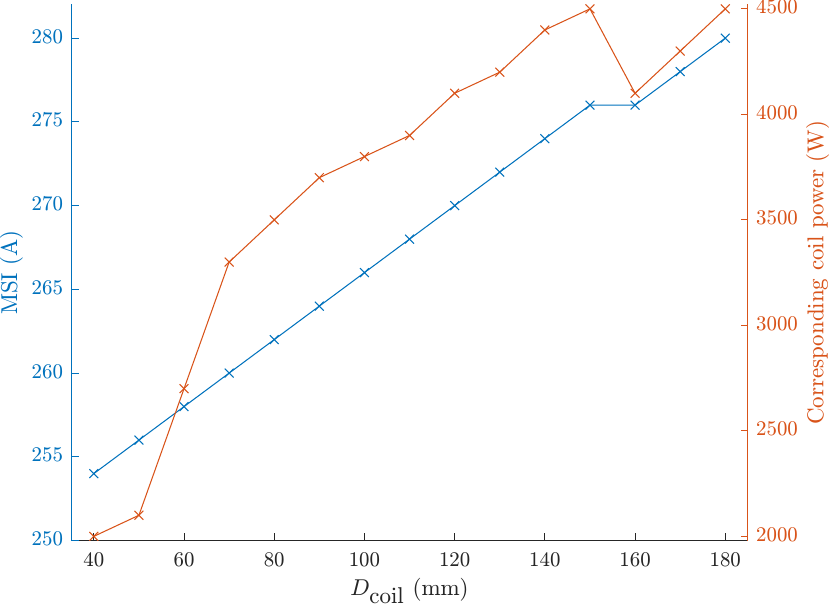}
	\caption{MSI of the MF coil for different coil distances, with $P_{\rm HF} = 3~kW$. The corresponding MF coil power is shown on the right axis.}
	\label{fig:MSI_Dcoil}
\end{figure}

\begin{figure}[!ht]
	\centering
	\includegraphics[width=0.7\linewidth]{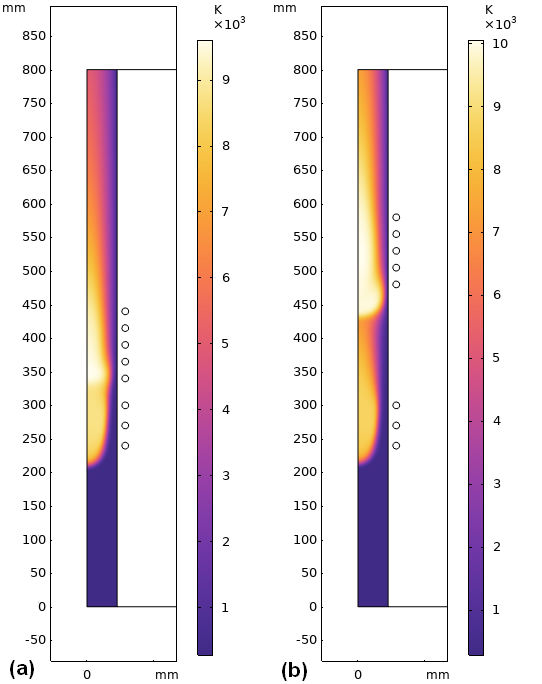}
	\caption{2D temperature plot when MSI is reached. \textbf{(a)} $D_{\rm coil} = 40~mm$ and \textbf{(b)} $D_{\rm coil} = 180~mm$.}
	\label{fig:T2D}
\end{figure}

In this subsection we present the evolution of the MSI for the MF coil when the coil distance is varied between $40~mm$ and $180~mm$. The excitation power of the HF coil is set to $P_{\rm HF} = 3~kW$. Outside of this distance range, the model was too instable to converge numerically. Figure~\ref{fig:MSI_Dcoil} shows that the MSI increases almost linearly with the coil distance. Indeed, the shorter the distance between the coils, the hotter the gas at the entrance of the MF coil, so less power is required to sustain the plasma, resulting in a drop of the MSI. This effect is clearly visible on the 2D temperature plot from figure~\ref{fig:T2D}. 

The corresponding MF coil power follows a more complex evolution, with a somewhat linear part between $70~mm$ and $150~mm$. It is interesting to note the important decrease in the MSI compared with the case where $P_{\rm HF} = 0~kW$: at equal coil distance, the required $444~A$ drop to $266~A$. While the corresponding coil power was about $15.2~kW$ in the first case, the total coil power in the latter case is $P_{\rm tot} = P_{\rm HF} + P_{\rm MF} = 3 + 3.8 = 6.8~kW$. Without considering losses (from cables, coil, etc.), the total power defined here relates directly to the ``wall plug'' power consumption. It shows that plasma operation requires less power with the dual frequency setup as compared with the single MF coil setup. Of course, the required power would be even less with a single HF coil setup, but the associated costs would also be higher. 

\newpage

\subsection{Varying the power of the HF coil}
\label{subsec:HFCoil}
\begin{figure}[!ht]
	\centering
	\includegraphics[width=0.7\linewidth]{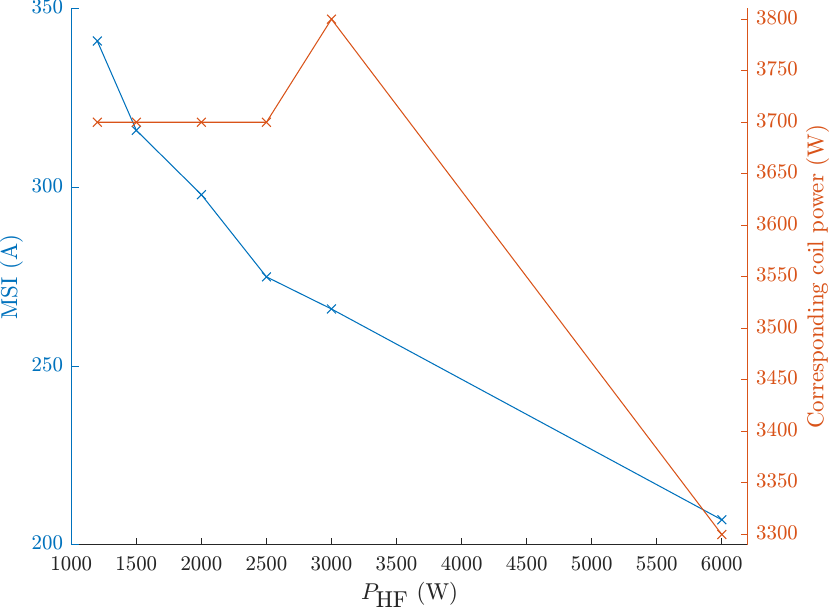}
	\caption{MSI of the MF coil for different power excitation of the HF coil, with $D_{\rm coil} = 100~mm$. The corresponding MF coil power is shown on the right axis.}
	\label{fig:MSI_PHF}
\end{figure}

The same type of study was realized with different values of $P_{\rm HF}$, while keeping the coil distance at $100~mm$. For the sake of completion, we included a simulation at $P_{\rm HF} = 6~kW$, but our interest lies in reducing the HF coil power rather than increasing it. Results are displayed in figure~\ref{fig:MSI_PHF}. As expected, the MSI of the MF coil increases when $P_{\rm HF}$ decreases. The influence of $P_{\rm HF}$ on the MSI is important: the MSI increases from $266~A$ at $P_{\rm HF} = 3~kW$ to $341~A$ at $P_{\rm HF} = 1.2~kW$, i.e., only $100~A$ below the MSI when the HF coil is off. Considering that the MF coil power remains almost constant while the required current for minimum operation increases rapidly with small decrease of $P_{\rm HF}$, it seems more reasonable to choose the HF power around $2.5~kW$ or $3~kW$ rather than around $1.2~kW$. Choosing a value larger than the MSP for pure argon also ensures that impurities from the walls or from leakage would not extinguish the plasma. Indeed, this atomic gas has a low ionization energy and does not undergo molecular dissociation, in contrast to $N_2$ or air.

\section{Other quantities of interest}
\label{subsec:profiles}

\subsection{Impedance of the MF coil}
\begin{figure}[!ht]
	\centering
	\includegraphics[width=0.8\linewidth]{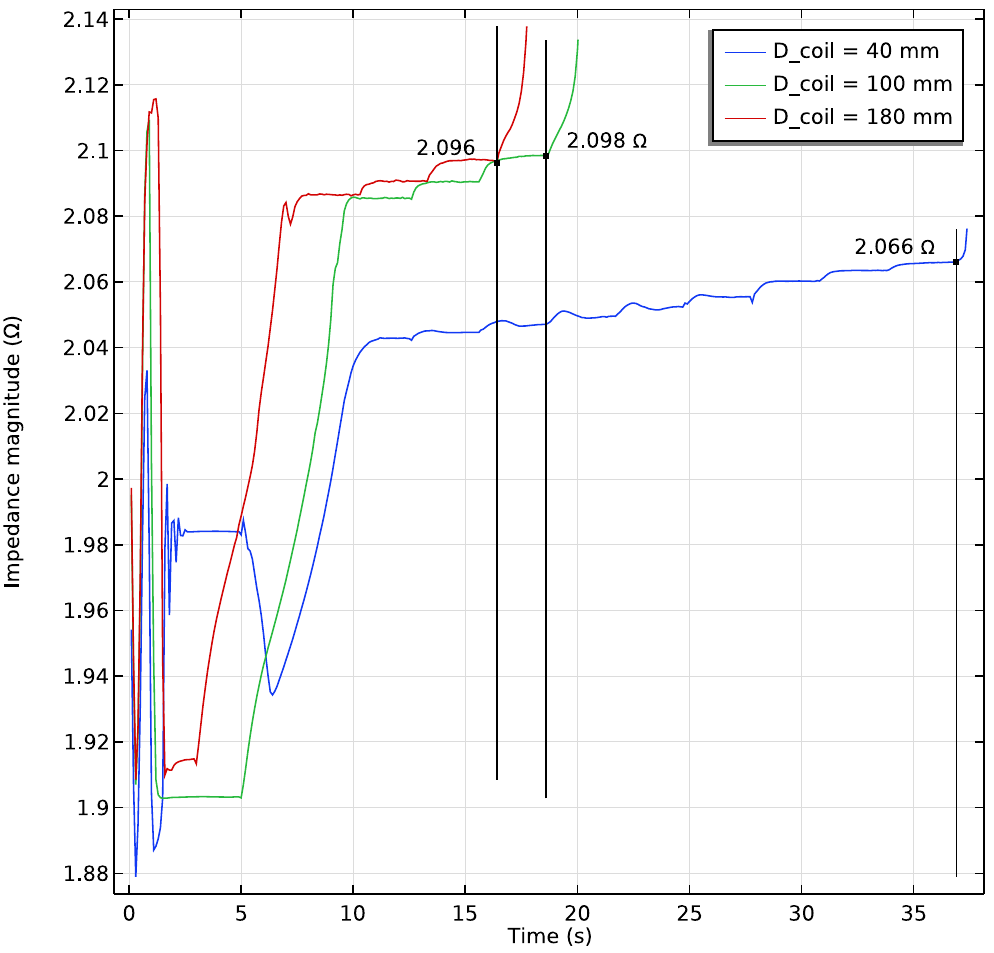}
	\caption{Magnitude of the impedance vs time for 3 different coil distances. The vertical black lines correspond to the time when the steady-state at MSI is reached.}
	\label{fig:impedanceMagnitude}
\end{figure}

\begin{figure}[!h]
	\centering
	\includegraphics[width=0.8\linewidth]{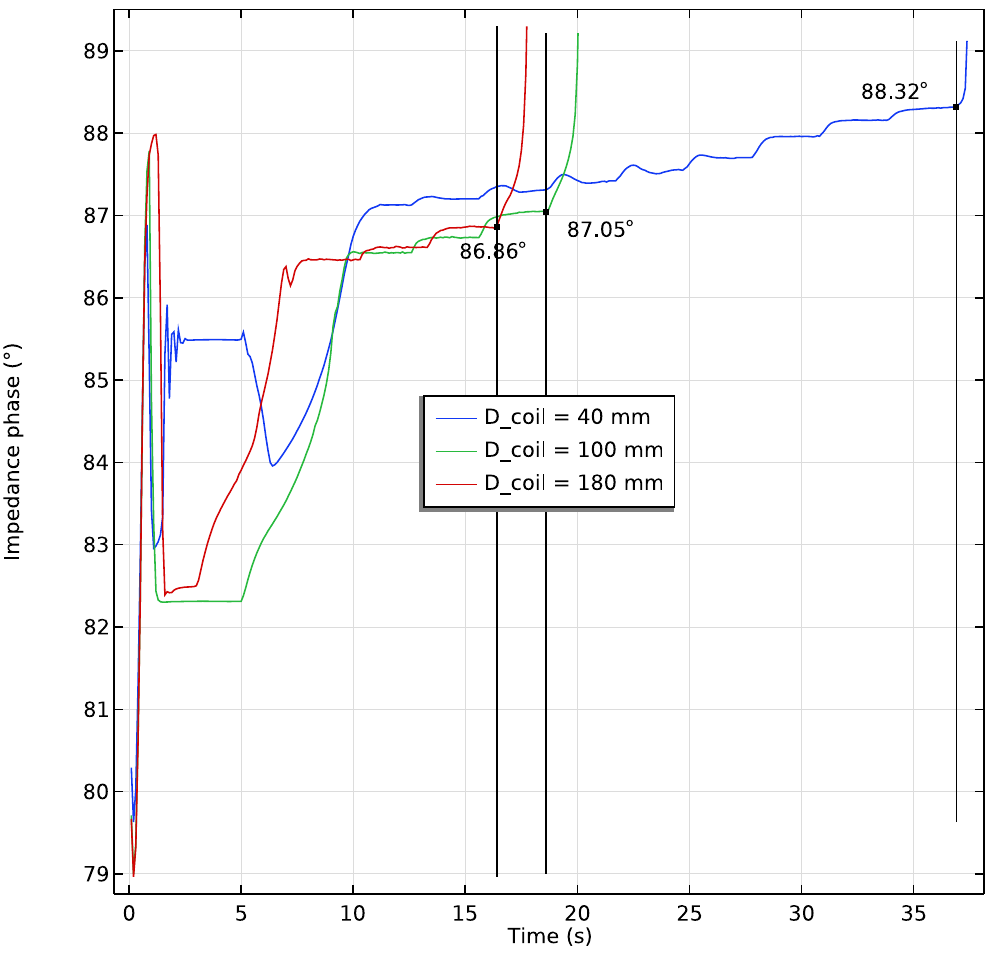}
	\caption{Phase of the impedance vs time for 3 different coil distances. The vertical black lines correspond to the time when the steady-state at MSI is reached.}
	\label{fig:impedancePhase}
\end{figure}

The magnitude and phase of the complex impedance of the MF coil are calculated in every simulation, at a time when the MSI is reached. Figure~\ref{fig:impedanceMagnitude} and figure~\ref{fig:impedancePhase} show 3 representative examples of the magnitude vs time and the phase vs time, respectively. Considering all the studies together, a maximum impedance magnitude of $2.121~\ohm$ is obtained for the case where $P_{\rm HF} = 1.2~kW$, while a minimum of $2.066~\ohm$ is obtained for $D_{\rm coil} = 180~mm$, i.e., a change of $2.66\%$ while choosing the minimum value as the reference. Similarly, a maximum impedance phase of $88.32\degree$ ($D_{\rm coil} = 40~mm$) and a minimum impedance phase of $86.86\degree$ ($D_{\rm coil} = 180~mm$) bring a change of $1.66\%$ overall. This means the allowed impedance operation region for a certain generator design is not considerably affected, and no further impedance matching is needed.

\subsection{Temperature and axial velocity profiles}
\begin{figure}[!h]
	\centering
	\includegraphics[width=0.7\linewidth]{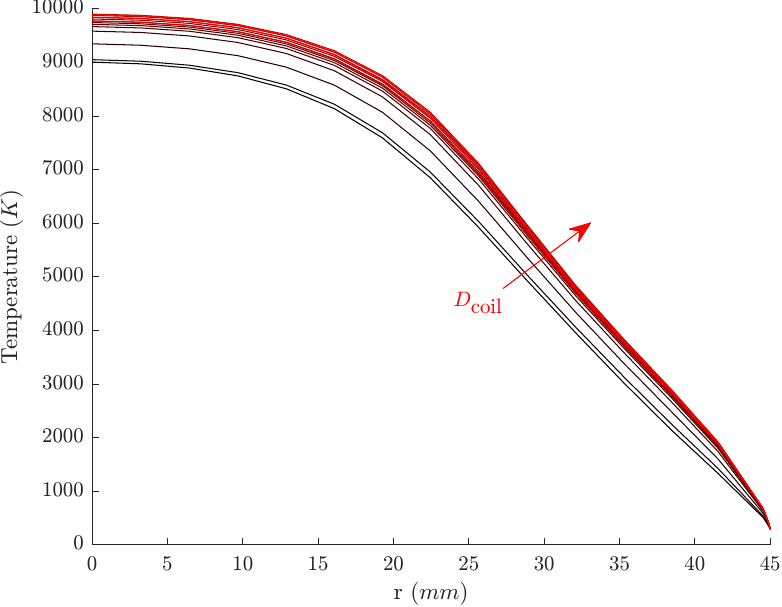}
	\caption{Radial temperature profiles $T(r,z_0)$ when the MSI is reached. The darker the color, the smaller $D_{\rm coil}$. $z_0$ is defined as the end of the MF coil.}
	\label{fig:T_vs_Dcoil}
\end{figure}

\begin{figure}[!htb]
	\centering
	\includegraphics[width=0.7\linewidth]{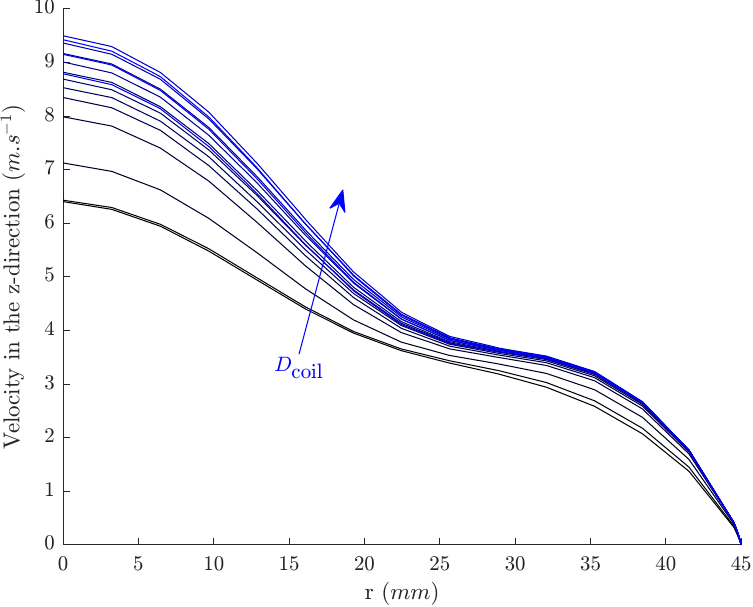}
	\caption{Radial longitudinal velocity profiles $V_z(r,z_0)$ when the MSI is reached. The darker the color, the smaller $D_{\rm coil}$. $z_0$ is defined as the end of the MF coil.}
	\label{fig:Vz_vs_Dcoil}
\end{figure}

\begin{figure}[!htb]
	\centering
	\includegraphics[width=0.7\linewidth]{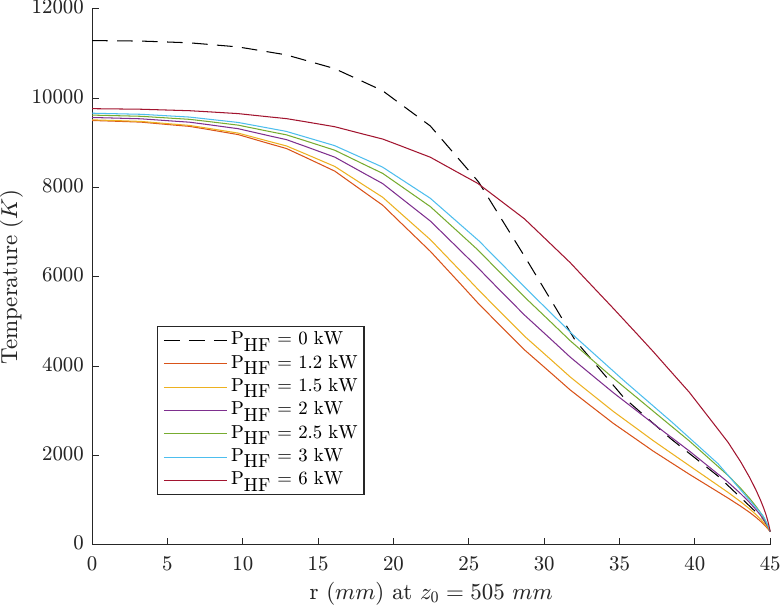}
	\caption{Radial temperature profiles $T(r,z_0)$ when the MSI is reached, for different HF coil powers. $z_0$ is defined as the end of the MF coil.}
	\label{fig:T_vs_PHF}
\end{figure}

\begin{figure}[!htb]
	\centering
	\includegraphics[width=0.7\linewidth]{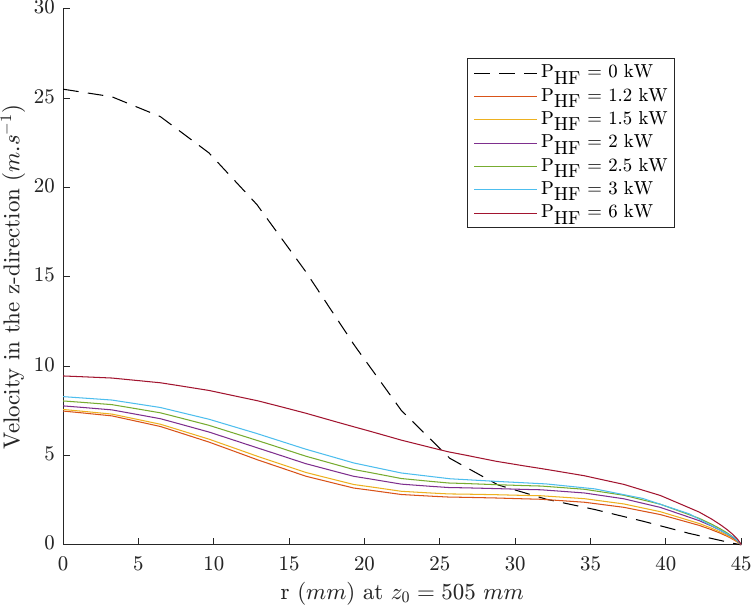}
	\caption{Radial longitudinal velocity profiles $V_z(r,z_0)$ when the MSI is reached, for different HF coil powers. $z_0$ is defined as the end of the MF coil.}
	\label{fig:Vz_vs_PHF}
\end{figure}

Although the steady-state reached at the MSI is much below the targeted total power of $1~MW$, it is interesting to investigate the temperature and velocity profiles along $r$ at a specific distance $z_0$. The calculated values represent the minimum of what is achievable while operating the dual frequency torch. Let us define $z_0$ as the end of the MF coil. With this definition, $z_0$ value is changing for the study where $D_{\rm coil}$ is varied, but fixed and equal to $505~mm$ for the study where $P_{\rm HF}$ is varied. Note that only the velocity in the $z$-direction (axial direction) is shown here, radial velocities having much lower values. Figure~\ref{fig:T_vs_Dcoil} and figure~\ref{fig:Vz_vs_Dcoil} show the temperature and velocity profiles when $D_{\rm coil}$ is varied. For all the temperature profiles, the maximum is located at $r = 0~mm$, which is a characteristic of a model without radiation. Interestingly, both the velocity and the temperature increase when $D_{\rm coil}$ increases, which is explained by the fact that the MF coil power is larger at high $D_{\rm coil}$ values. While the MF coil power (hence the power coupled into the plasma) at $D_{\rm coil} = 180~mm$ is twice more than the one at $D_{\rm coil} = 40~mm$, the increase in temperature is only $\sim1000~K$. The influence on the velocities seems relatively larger in comparison. Such considerations will prove useful when optimizing the torch for heating applications. For the same reasons, the temperature and the velocity increase when $P_{\rm HF}$ increases, as shown in figure~\ref{fig:T_vs_PHF} and figure~\ref{fig:Vz_vs_PHF}.

\subsection{Conductive and convective heat flows at the end of the MF coil}
\begin{figure}
	\centering
	\includegraphics[width=0.7\linewidth]{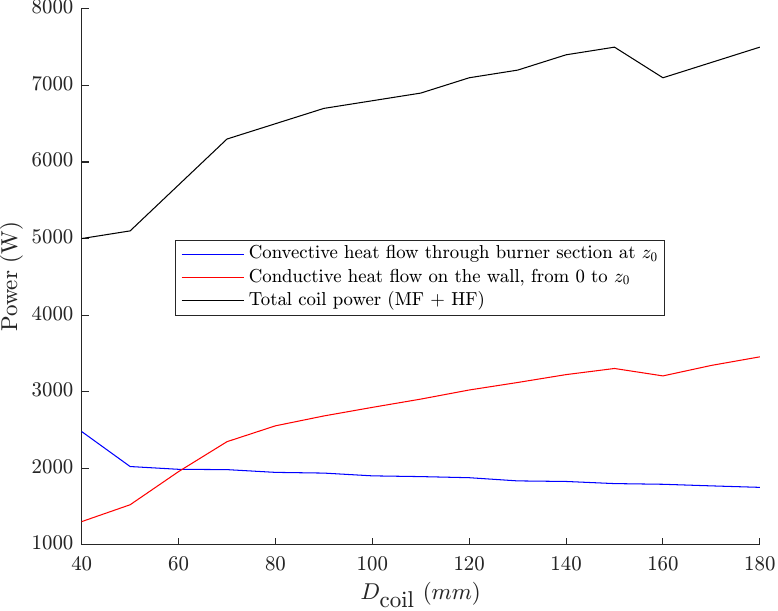}
	\caption{Convective heat flow over the burner section at $z_0$ and conductive heat flow over the burner wall up to $z_0$. }
	\label{fig:heat}
\end{figure}

Lastly, we calculate the convective heat flux integrated over a disk of radius $R_{\rm burner}$, located at $z_0$, and the conductive heat flux integrated over the burner wall on a cylindrical surface of length $z_0$. Results are plotted in figure~\ref{fig:heat} vs the coil distance $D_{\rm coil}$ and compared with the total coil power. Here we choose to show only the case where $D_{\rm coil}$ is varying, so that $z_0$ also changes but $P_{\rm HF}$ is fixed at $3~kW$. The sum of both heat flows is lower than the total coil power, mainly because the conductive heat in the axial direction at $z_0$ is not shown here (and to a lesser extent because of coil losses). On the one hand, the convective heat flow is almost constant vs $D_{\rm coil}$, even though $P_{\rm MF}$ is increasing. On the other hand, the conductive heat to the burner wall is increasing in a similar way as the coil power, which is expected considering that the surface of the burner wall exposed to plasma is increasing with $D_{\rm coil}$. However, it also means that a larger gap between the two coils is reducing the efficiency of the torch in terms of convected heat at the outlet. On top of this, if the conducted heat at the wall is too high, one risks to damage the torch by melting it. While here we present integrated results including parts where the torch is cold (at low $z$ values), the typical profile of the conductive heat flux is extremely non-linear and exhibits very strong peaks close to the coils. These peak values can be used to define a maximum operable range depending on how resistant the wall material is.

\subsection{Grid-independence study}
\label{subsec:gridinpendence}

\begin{figure}[!ht]
	\centering
	\includegraphics[width=1\linewidth]{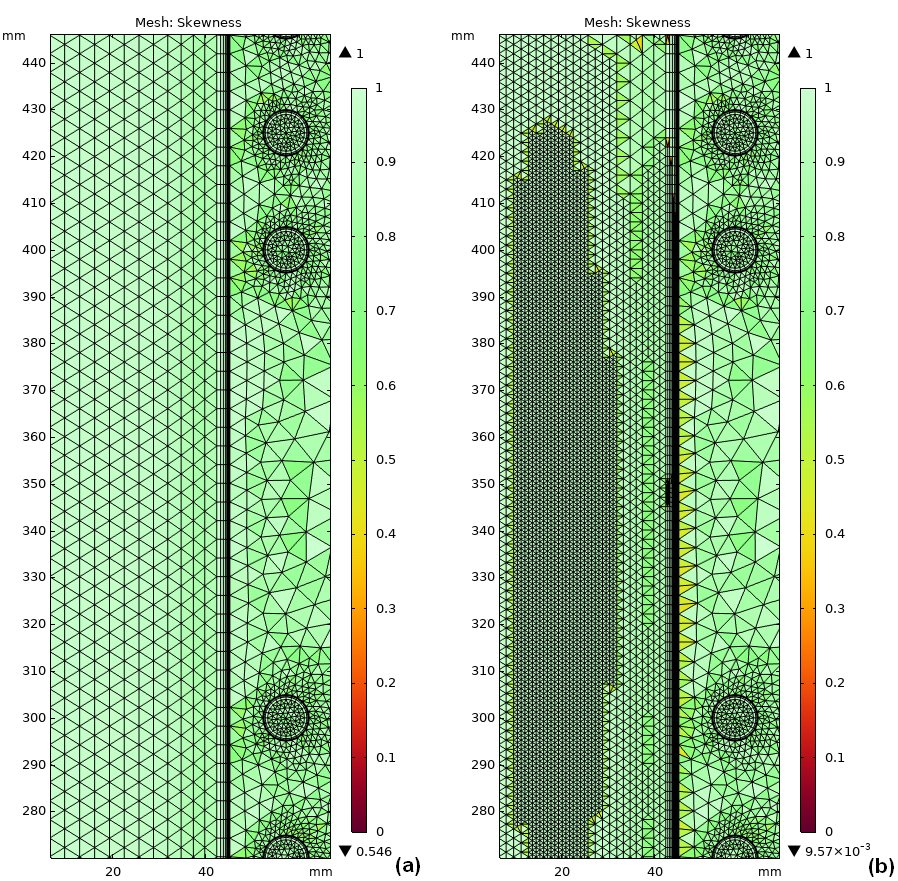}
	\caption{Comparison of the plasma region grids between: (a) the default mesh and (b) the refined mesh.}
	\label{fig:mesh_comparison}
\end{figure}

The default mesh defined in subsection~\ref{subsec:meshconstruction} was refined to study whether or not the results depend strongly on the grid size. Only the default case (table~\ref{tab:defaultValues}) up to $6~s$ is evaluated with the new grid, hence corresponding to a stabilized plasma at full power without considering the ramping down of the coil current. The refined mesh was generated automatically from the default one with the ``adaptive mesh refinement'' feature from COMSOL. It consists in 49,753 elements, so about $2.7$ times more than the default mesh, all refined cells being located in the plasma region. Figure~\ref{fig:mesh_comparison} shows a grid zoom of the plasma region for both meshes. Its average element quality is $0.9109$. The computational time is $3$ times longer than with the default mesh. 

\begin{figure}[!ht]
	\centering
	\includegraphics[width=0.7\linewidth]{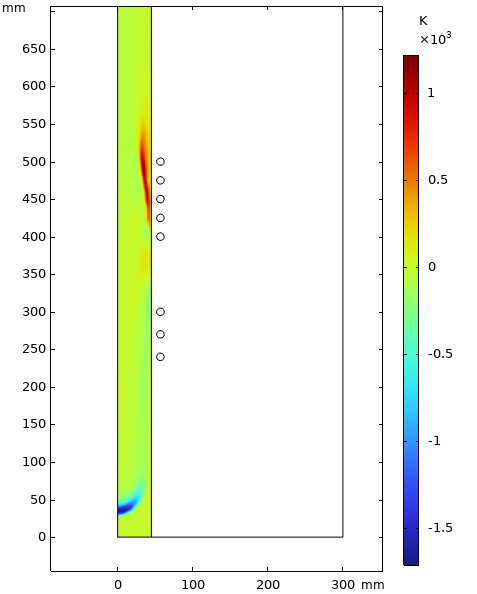}
	\caption{Difference of temperature distribution between the default mesh and the refined mesh. A positive difference means that the default mesh temperature is higher.}
	\label{fig:temperaure_difference}
\end{figure}

Both the magnitude and the phase of the impedance change by $0.02\%$ at $t = 6~s$, which is negligible. The power coupled into the plasma from both coils amounts to $34.68~kW$ for the default mesh, while it is $34.02~kW$ for the refined mesh, hence a difference of $1.9\%$. This is likely due to a slightly different temperature distribution within the plasma, as shown in figure~\ref{fig:temperaure_difference}. Nonetheless, the discrepancies are small enough to confirm that the results are not too grid-dependent. 

\section{Conclusion}
Using COMSOL Multiphysics\textsuperscript{\textregistered}, we created a numerical model of an ICP with two different frequencies. We confirmed the well-known fact that less coil power is required to ignite and sustain a plasma when the RF current frequency increases. More interestingly, we quantified the minimum sustaining current when the coil distance and the power of the high-frequency coil are changed. The great drop of the MSI when a high-frequency assistance is provided offers a large range of possible current values to optimize the torch. It further helps ignition and operation of the plasma, which is difficult with a medium-frequency coil alone. In parallel, the HF coil does not need high power values and the targeted total power of $1~MW$ can be mainly provided by the MF coil. Within the context of this work, the operation of a HF coil and the distance between coils do not affect considerably the impedance seen at the MF coil. This implies that, for a certain power generator with its corresponding impedance matching, their design does not need additional adaptation for their usage in a dual-frequency system.

Although planned to be the first step towards a more complex model of a dual frequency torch, this work will not be continued by our group in the near future. This is due to a change of main focus, as well as a lack of facilities in the laboratory. Past experiments confirmed the feasibility of a RF-RF design, but the measurements are too scarce to be presented here. The corresponding equipment being used for another project, we do not expect any up-to-date dataset soon to validate the model. Initially, we intended to include more gases to our model, like air, $O_2$ and $CO_2$, with a specific focus on the widely used $N_2$. Radiation is also a crucial heat exchange process when considering whether the torch wall can sustain operation or not. However, an accurate radiation modeling is difficult, as it heavily depends on global assumptions of transparency/opacity of the plasma, which is not homogeneous. The feed gas must also be well-known, as the main parameter would be the total volumetric emission coefficient, described as temperature-dependent on COMSOL. 

On the other hand, 3D simulations are not considered as a viable route yet, because of the increasing complexity and computational cost. Even very simplified 3D geometries designed for test cases could not converge when we tested them, while taking more than a week to compute the first seconds of the simulation. As of now, a 2D axisymmetric model is the only affordable way for parametric studies such as the ones presented here. 

Geometry is definitely another aspect deserving more investigation. The design presented here is the simplest possible for a dual frequency torch, which allows us to characterize key quantities with a minimum of influencing factors. Most typical ICPs include different inlets, with long sheath walls for example. The gap between the wall and the coil has an impact on the coupling efficiency of the magnetic field. Modifying the torch radius (e.g., smaller for the HF coil) or using nozzles can prevent some of the plasma from going upstream and change the overall flow pattern. Lastly, vortexes trapping heat in non-suitable locations can appear at higher coil currents, sometimes completely changing the region where the power is coupled into the plasma. A detailed geometric study becomes therefore mandatory when one wants to optimize the torch for realistic operating values.



\bibliographystyle{aomalpha}
\bibliography{biblio}
\addcontentsline{toc}{section}{References}

\end{document}